\documentstyle[12pt,aaspp4]{article}
\newcommand{\pad}{\phantom{$\pm$0.0}}

\begin{document}

\title{The Large-scale $J$ = 3$\rightarrow$2 and $J$ = 2$\rightarrow$1 CO
Emission from M17 \\
and its Implications for Extragalactic CO Observations}
\author{Christine D. Wilson\altaffilmark{1}}
\author{John E. Howe\altaffilmark{2}}
\author{Michael L. Balogh \altaffilmark{3}}
\bigskip

\altaffiltext{1}{Department of Physics and Astronomy, McMaster University,
Hamilton, Ontario L8S 4M1 Canada} 
\altaffiltext{2}{Five College Radio Astronomy Observatory, Department
of Physics and Astronomy, University of Massachusetts, Amherst, MA 01003}
\altaffiltext{3}{Department of Physics and Astronomy, University of 
Victoria, Victoria, BC V8W 3P6 Canada}

\begin{abstract}

We observed a $10 \times 20$ pc region of the molecular cloud M17 in
the $^{12}$CO and $^{13}$CO $J$ = 3$\rightarrow$2 and $J$ =
2$\rightarrow$1 transitions at the James Clerk Maxwell Telescope to
determine their global behavior and to assess the reliability of using
ratios of CO line intensities integrated over an entire cloud to
determine the physical conditions within the cloud.  Both the
$^{12}$CO/$^{13}$CO $J$ = 2$\rightarrow$1 and $J$ = 3$\rightarrow$2
line ratios correlate with the $^{13}$CO integrated intensity, with
smaller line ratios observed at locations with large integrated
intensities. This correlation is likely due to variations in the
column density from one position to another within M17.
The $^{12}$CO and
$^{13}$CO ($J$ = 3$\rightarrow$2)/($J$ = 2$\rightarrow$1) line ratios
show no significant variation from place to place within M17, even on
the peak of the photon-dominated region.
A Large Velocity Gradient analysis of globally averaged line
ratios gives a kinetic temperature of 30~K, a density of $0.2-5\times 10^5$
cm$^{-3}$, and a $^{12}$CO column density of $1\times 10^{18}$
cm$^{-2}$. This result is in reasonable agreement with the results
obtained for individual lines-of-sight through the cloud, which
suggests that the typical physical conditions in a molecular cloud can
be determined using CO line ratios integrated over the entire cloud.

We estimate the global CO line ratios for M17 to be
$^{12}$CO/$^{13}$CO $J$ = 2$\rightarrow$1: $4.5\pm0.7$;
$^{12}$CO/$^{13}$CO $J$ = 3$\rightarrow$2: $3.7\pm0.9$;
$^{12}$CO ($J$ = 3$\rightarrow$2)/($J$ = 2$\rightarrow$1): $0.76\pm0.19$;
$^{13}$CO ($J$ = 3$\rightarrow$2)/($J$ = 2$\rightarrow$1): $1.3\pm0.3$.
These line ratios generally agree quite well with previous measurements
of individual Galactic molecular clouds.  There appears to be a clear
trend of increasing $^{12}$CO/$^{13}$CO $J$ = 2$\rightarrow$1 and $J$ =
3$\rightarrow$2 line ratios as one moves from Galactic molecular cloud
cores to entire Galactic molecular clouds to normal galaxies,
 similar to the trend seen previously for the
$^{12}$CO/$^{13}$CO $J$ = 1$\rightarrow$0 line. These new observations
of M17 show that the difference between the $^{12}$CO/$^{13}$CO line
ratios for Galactic molecular clouds and the disks of spiral galaxies
occurs for all three of the lowest rotational transitions. The most
likely explanation of the high line ratios for normal galaxies is a
significant contribution to the CO emission by low column density
material, such as diffuse molecular clouds or the outer envelopes of
giant molecular clouds. Radial gradients in the relative contribution
of low and high column density material in galaxies could be a
significant source of uncertainty in derivations of the physical
properties of molecular gas in external galaxies.

\end{abstract}


\section{Introduction}

Because molecular clouds contain structure over many different physical
scales, a full understanding of the properties of molecular clouds
requires not only detailed observations at high spatial resolution but
also large-area maps of entire clouds.  Information on both the large-
and small-scale properties of molecular clouds is now quite commonly
available from observations of the $J$ = 1$\rightarrow$0 transitions of
the isotopomers of CO.  These data can be used to determine cloud
masses and column densities, as well as to study the small-scale clumpy
structure of the clouds. However, to obtain more than a crude estimate
of the temperature and density structure of molecular clouds requires
similar observations in the higher rotational transitions of these
molecules (e.g., \markcite{c90}Castets et al.\ 1990).  With the advent
of sensitive submillimeter telescopes, such data are becoming more
common. However, because of the small beam of most submillimeter
telescopes compared to the sizes of molecular clouds, such observations
are usually limited to single positions or small maps of particularly
interesting regions within a cloud, such as photon-dominated regions
or star-forming cores (e.g., \markcite{sg90}Stutzki \& G\"usten 1990).
Thus, although these studies can reveal important details in the
small-scale structure of molecular clouds, information on the
large-scale properties of molecular clouds in these higher rotational
transitions is generally lacking.  Some exceptions are the Orion
molecular cloud, which has been completely surveyed in the $^{12}$CO
$J$ = 2$\rightarrow$1 transition with 9\arcmin\ resolution using a
specially-designed telescope (\markcite{s94}Sakamoto et al.\ 1994),
the Orion B ($^{12}$CO and $^{13}$CO $J$ = 2$\rightarrow$1) and
S140 ($^{13}$CO $J$ = 2$\rightarrow$1) molecular clouds, mapped at
3\arcmin\ resolution using re-imaging optics on the Caltech Submillimeter
Observatory 10.4 m telescope (\markcite{pj94}Plume \& Jaffe 1994;
\markcite{pjk94}Plume, Jaffe, \& Keene 1994), and the Orion B
and Rosette molecular clouds, mapped at 2\arcmin\ resolution with
the KOSMA telescope (\markcite{k96}Kramer, Stutzki, \& Winnewisser
1996; \markcite{S98}Schneider et al.\ 1998).

Large-scale maps of Galactic molecular clouds in the higher rotational
transitions of CO can be used to answer many questions concerning the
global properties of molecular clouds. For example, variations in the
CO emission line ratios can provide clues to the existence and nature
of large-scale density and temperature gradients in molecular clouds.
In the Orion molecular cloud, the $^{12}$CO ($J$ =
2$\rightarrow$1)/($J$ = 1$\rightarrow$0) line ratio varies from about 1
in the central ridge to about 0.5 near the edges of the cloud
(\markcite{s94}Sakamoto et al.\ 1994).  Sakamoto et al.\ argue that
this line ratio trend, which also causes the Orion molecular cloud to
appear smaller in the $^{12}$CO $J$ = 2$\rightarrow$1 transition, is
due to a ten-fold decrease in the gas density from the core to the
periphery of the cloud.  Furthermore, there are hints that the cloud
appears smaller still in the $^{13}$CO $J$ = 2$\rightarrow$1 transition
(\markcite{s94}Sakamoto et al.\ 1994).  Thus, we might expect line
ratio variations in even higher rotational transitions of CO, since
their upper levels are more likely to be subthermally excited over a
larger region of a molecular cloud than the $J = 1$ or $J = 2$ levels.

Large-scale maps of Galactic molecular clouds can also provide crucial
information for interpreting the many recent observations of these
transitions in external galaxies, in which emission line ratios are
used to determine the average density and temperature of the molecular
cloud ensemble. One implicit assumption of such measurements, that the
physical conditions inside molecular clouds do not change much from one
cloud to another, has recently received observational support from a
multi-transition study of individual molecular clouds in the nearby
galaxy M33 (\markcite{w97}Wilson, Walker, \& Thornley 1997).  However,
a second implicit assumption of all extragalactic studies (including
the M33 study), that the different rotational transitions are produced
by the same gas component of the clouds and hence have a similar
angular extent, is called into question by the \markcite{s94}Sakamoto
et al.\ (1994) Orion observations, and should be tested by large-area
observations of other Galactic molecular clouds. In addition, a
comparison of the density and temperature derived for individual lines
of sight through a cloud with those derived from the cloud viewed as a
whole can show whether global CO line ratios may be used to provide a
reliable measure of the physical properties of molecular clouds.

In this paper, we present data over a $10 \times 20$ pc area of the
molecular cloud M17 for the $^{12}$CO and $^{13}$CO $J$ =
3$\rightarrow$2 and $J$ = 2$\rightarrow$1 lines.  At a distance of 2.2
kpc (\markcite{c80}Chini, Els\"asser, \& Neckel 1980), M17 is
representative of the kinds of star-forming giant molecular clouds that
are thought to emit most of the low-J CO emission in galaxies (e.g.,
\markcite{c85}Crawford et al.\ 1985; \markcite{w89}Wolfire et
al.\ 1989).  M17 has been the subject of numerous studies since the
original CO $J$ = 1$\rightarrow$0 map of \markcite{l76}Lada (1976)
outlined the northern and southern pieces of the nearby molecular
cloud, which is the brightest and most massive fragment of a molecular
cloud complex extending southwest nearly 170 pc along the Sagittarius
spiral arm (\markcite{e79}Elmegreen, Lada, \& Dickinson 1979).
However, most of these studies have concentrated on the
photon-dominated region near the boundary of the HII region and the
M17SW molecular cloud (e.g., \markcite{m92}Meixner et al.\ 1992;
\markcite{g92}Greaves, White, \& Williams 1992; \markcite{sg90}Stutzki
\& G\"usten 1990; \markcite{s88}Stutzki et al.\ 1988;
\markcite{r87}Rainey et al.\ 1987).  Our study instead attempts to
provide a view of the entire cloud, including regions quite distant
from the photon-dominated region.  Although our study does not cover
the entire extended molecular cloud complex, it should provide us with
an unbiased view of the extended bright emission region of the M17
molecular cloud.  We adopt the following nomenclature in this paper.
M17 refers to the entire cloud (\markcite{l76}Lada 1976), while M17
North and M17 South refer to the northern and southern halves of the
cloud.  M17 SW, or simply ``the PDR'' refers to the peak emission from
the photon-dominated region (PDR) located in M17 South.  The
observations and data reduction are discussed in \S 2.  The CO
integrated intensities and line ratios are summarized in \S 3 (with
details of the line ratio calculations given in the Appendix), and
solutions for the physical conditions (density, temperature, and column
density) for five locations around the cloud as well as for the entire
cloud are presented in \S 4.  A comparison with extragalactic data and
the implications for determining physical conditions in external
galaxies are presented in \S 5.  The paper is summarized in \S 6.

\section{Observations and Data Reduction}

We observed the giant molecular cloud M17 with the James Clerk Maxwell
Telescope (JCMT) in the $^{12}$CO and $^{13}$CO $J$ = 2$\rightarrow$1
and $J$ = 3$\rightarrow$2 rotational transitions on 1993 August 25--28,
using the Dwingeloo Autocorrelation Spectrometer as the backend.  The
half-power beamwidth of the JCMT is 22\arcsec\ at 230 GHz and
15\arcsec\ at 345 GHz.  The map center is
($\alpha$(1950) = 18$^{\rm h}$17$^{\rm m}$21\fs0,
$\delta$(1950) = $-$16\arcdeg06\arcmin45\arcsec),
roughly the center of the bright CO emission from
the M17 molecular complex (\markcite{l76}Lada 1976).  We observed in
position switching mode with a reference position 35\arcmin\ east of
the map center. Although this location is far enough away that there is
no emission over the velocity range of the emission from M17 ($\sim
10-30$ km s$^{-1}$), the reference position does contain emission
around $\sim 50-80$ km s$^{-1}$ that produces negative dips in the
spectra.  We observed the peak of the PDR in M17 South $(\Delta\alpha,
\Delta\delta) = (2\arcmin,-6\arcmin)$ every 30 to 60 minutes as a check
on the stability of the calibration.  From the repeatability of the PDR
position, we estimate the measurement uncertainties in the line ratios to
be $\pm 4\%$ for $^{12}$CO/$^{13}$CO $J$ = 2$\rightarrow$1, $\pm 10\%$ for
$^{12}$CO/$^{13}$CO $J$ = 3$\rightarrow$2, $\pm 9\%$ for $^{12}$CO ($J$
= 3$\rightarrow$2)/($J$ = 2$\rightarrow$1), and $\pm 7\%$ for $^{13}$CO
($J$ = 3$\rightarrow$2)/($J$ = 2$\rightarrow$1). The true uncertainty
in the ($J$ = 3$\rightarrow$2)/($J$ = 2$\rightarrow$1) line ratios is
probably considerably larger because of the need to convolve the data
to match the different beam sizes, while the $^{12}$CO/$^{13}$CO line
ratios are probably somewhat larger because the relative weakness of
the $^{13}$CO lines makes baseline uncertainties and crowding by weak
features due to other clouds more significant. Thus, we adopt absolute
uncertainties of $\pm 15\%$ for the $^{12}$CO/$^{13}$CO line ratios and
$\pm 25\%$ for the ($J$ = 3$\rightarrow$2)/($J$ = 2$\rightarrow$1) line
ratios.

We obtained three different data sets with different mapping strategies
during the run. The first data set consists of a map of the entire
region mapped by \markcite{l76}Lada (1976) ($16^\prime \times
32^\prime$, corresponding to $10.2 \times 20.4$ pc) in the $^{12}$CO
and $^{13}$CO $J$ = 2$\rightarrow$1 lines with 2\arcmin\ sampling. The
second data set consists of three cuts in Right Ascension and one cut
in Declination through the cloud in the $^{12}$CO and $^{13}$CO $J$ =
3$\rightarrow$2 lines. The Right Ascension cuts are located at
$\Delta\delta = -8\arcmin,0\arcmin,8\arcmin$ and have
2\arcmin\ sampling in right ascension, while the Declination cut is
located at $\Delta\alpha = 2\arcmin$ and has
2\arcmin--6\arcmin\ sampling in declination. The third data set
consists of observations of all four transitions at nine locations
around the cloud, five in the Declination cut with $6^\prime$ sampling
and four chosen to sample different regions of the cloud.  For this
data set, we mapped the $^{12}$CO and $^{13}$CO $J$ = 3$\rightarrow$2
emission on a three by three point grid with 8\arcsec\ sampling to
enable us to convolve the data to the larger beam of the $J$ =
2$\rightarrow$1 data.

As a check that the highly undersampled $J$ = 2$\rightarrow$1 maps provide
a reasonable overview of the structure of M17, we also obtained a half-beam
sampled $^{13}$CO $J$ = 1$\rightarrow$0 map on 1997 June 11--12 with the
14~m telescope of the Five College Radio Astronomy Observatory (FCRAO),
using the 15-element array receiver QUARRY (\markcite{e92}Erickson et
al.\ 1992) and the backend autocorrelation spectrometers set to a spectral
resolution of 0.26 km s$^{-1}$.  The half-power beamwidth of the FCRAO
observations is 47\arcsec.

M17 has a very complicated structure that includes clumps similar to
the beam size as well as emission extended over about half a degree
(\markcite{l76}Lada 1976; \markcite{sg90}Stutzki \& G\"usten 1990).
Given the different angular scales over which emission is seen, we
adopt a temperature scale which is an average of the main beam and
$T_R^*$ temperature scales, and should provide the best approximation
to the radiation temperature of the emission.  For the 220--230 GHz
JCMT observations, the forward scattering and spillover efficiency,
$\eta_{fss}$, is 0.80 and the main beam efficiency, $\eta_{MB}$, is
0.69, while at 330--345 GHz $\eta_{fss} = 0.70$ and $\eta_{MB} =
0.58$.  Thus, we adopt $\eta = (0.5/\eta_{fss} + 0.5/\eta_{MB})^{-1} =
0.76$ for the $J$ = 2$\rightarrow$1 data and $\eta = 0.60$ for the $J$
= 3$\rightarrow$2 data, so that for a given transition $T_R =
T_A^*/\eta$.  All data in this paper are given in this hybrid
temperature scale.  The 110 GHz FCRAO beam efficiencies are
$\eta_{fss} = 0.72$ and $\eta_{MB} = 0.52$, giving a hybrid
efficiency $\eta = 0.60$.

The JCMT data were reduced using the Bell Labs data reduction package COMB.
The spectra were binned to a resolution of 0.213 km s$^{-1}$ (the
intrinsic resolution of the $^{13}$CO $J$ = 2$\rightarrow$1 spectra)
and first to (occasionally) third order polynomial baselines removed.
For the nine locations with $J$ = 3$\rightarrow$2 maps, each set of
nine $J$ = 3$\rightarrow$2 spectra was convolved to simulate a
22\arcsec\ beam.  Integrated intensities were measured for each
spectrum using a velocity range appropriate to isolate emission from
the M17 cloud, usually some subset of the range from 10 to 30 km
s$^{-1}$. For many of the spectra, emission lines appear at larger and
occasionally smaller velocities, but this emission was not included in
the analysis. The emission peaks are clearly separated from the M17
emission lines and probably represent separate clouds along the same
line of sight.  The FCRAO data were reduced in the same manner using
the SPA data reduction package, except the spectra were not re-binned
and the velocity range for the integrated intensity was the same for
each position.

\section{Observational Results}

We present the integrated intensities and line ratios for the
individual positions in Tables~\ref{tbl-1}-\ref{tbl-4}.  All the line
ratios in Tables~\ref{tbl-1}-\ref{tbl-4} are ratios of integrated
intensities, rather than ratios of the peak brightness temperatures.
Tables~\ref{tbl-1} and~\ref{tbl-2} show the $^{12}$CO $J$ =
2$\rightarrow$1 integrated intensities and the $^{12}$CO/$^{13}$CO $J$
= 2$\rightarrow$1 line ratios.  Table~\ref{tbl-3} gives the $J$ =
3$\rightarrow$2 data for the four cuts through the cloud, while
Table~\ref{tbl-4} gives the results for all four transitions for the
nine positions with beam-matched observations.  In Table~\ref{tbl-5} we
give estimates of the line ratios appropriate for the M17 cloud as a
whole and 
compare our measured line ratios with previous Galactic measurements;
in general, the agreement with previous work is quite good.
We describe the calculations of the line ratios and
beam-matched $J$ = 3$\rightarrow$2 line intensities in the Appendix.

Contour plots of the $^{12}$CO and $^{13}$CO $J$ = 2$\rightarrow$1
integrated intensity are shown in Figure~\ref{fig-1}a and
Figure~\ref{fig-1}b.  The $^{13}$CO $J$ = 1$\rightarrow$0 integrated
intensity map is shown in Figure~\ref{fig-1}c. The similarity of the
$J$ = 2$\rightarrow$1 maps to the $J$ = 1$\rightarrow$0 map suggests
that the undersampled $J$ = 2$\rightarrow$1 maps provide a reasonably
unbiased measurement of the large-scale structure of M17.

Inspection of the $^{12}$CO/$^{13}$CO $J$ = 2$\rightarrow$1 line ratios
given in Table~\ref{tbl-2} shows that lower values of the line ratio
($<4$) are found around the PDR M17 SW, along a ridge running to the
north-west of the PDR, and in a small region in M17 North. In general,
the regions with lower line ratios tend to have higher integrated
intensities. In addition, larger values of the line ratio ($>8$) are
generally found on the edges of the cloud.  However, not all the edge
locations have high line ratios and, conversely, not all locations with
high line ratios are on the edge of the cloud.

The $J$ = 3$\rightarrow$2 data for the cuts across the cloud given in
Table~\ref{tbl-3} show similar trends to those seen in the $J$ =
2$\rightarrow$1 data, with lower $^{12}$CO/$^{13}$CO $J$ =
3$\rightarrow$2 line ratios near the PDR M17 SW and near the peak
emission in the cloud M17 North, and higher ratios towards the edges of
the cloud (Figure~\ref{fig-2}).  (The high line ratios near the center
of the northern cut are due to the relatively weak integrated
intensities found in the middle of this cut (see Figure~\ref{fig-1}).)
Fourteen of the observed positions show two or more peaks in one or
both of the spectra. Some examples of these spectra are shown in
Figure~\ref{fig-3}.  For about half of these positions the
$^{12}$CO/$^{13}$CO $J$ = 3$\rightarrow$2 line ratios in the different
components agree to within 20\%, while for the remaining positions the
line ratios in the different components can differ by more than a
factor of three. At a few of the positions, the $^{12}$CO $J$ =
3$\rightarrow$2 line appears to be self-absorbed, since the dip in the
$^{12}$CO spectrum occurs near the velocity of the single peak of the
$^{13}$CO $J$ = 3$\rightarrow$2 spectrum.

\section{CO Line Ratios and Physical Conditions in M17}

\subsection{The Relationship Between Line Ratios and Physical Conditions}

Before we discuss the physical conditions in M17 implied by the
observations, we briefly outline the behavior of CO line ratios as the
physical parameters of the gas vary.  The $J = 2$ and $J = 3$
rotational levels of CO lie 17~K and 33~K above the ground rotational
state, and in the absence of photon trapping, the $J$ = 2$\rightarrow$1
and $J$ = 3$\rightarrow$2 transitions thermalize at an $\rm H_2$
density of $4 \times 10^4$ cm$^{-3}$ and $1.6 \times 10^5$ cm$^{-3}$,
respectively.

In a molecular cloud with constant abundances, the emergent line
intensities depend on the density, temperature, column density, and
velocity dispersion of the gas.  We modeled the behavior of the CO line
intensities as a function of $\rm H_2$ density, kinetic temperature,
and column density per unit velocity dispersion interval using the
Large Velocity Gradient (LVG) approximation (see \S 4.2).  In
Figure~\ref{fig-4} we show the variation of the $^{12}$CO and $^{13}$CO
line ratios with all but one parameter fixed at a value appropriate for
a moderately warm, dense molecular cloud:  $T_K = 30$~K, $n(\rm H_2) =
10^5$ cm$^{-3}$, and $N({\rm CO})/\Delta V = 5\times10^{17}$ cm$^{-2}$ (km
s$^{-1}$)$^{-1}$.  We assume a constant [$^{12}$CO]/[$^{13}$CO]
abundance ratio of 50 (see \S 4.2).  In all three plots, the ratios are
of peak line radiation temperature $T_R$, and the left-hand scale
refers to the ($J$ = 3$\rightarrow$2)/ ($J$ = 2$\rightarrow$1) ratios
(solid lines), while the right-hand scale refers to the
$^{12}$CO/$^{13}$CO ratios (dotted lines).  The behavior of ratios of
integrated line intensities is qualitatively similar, with minor
differences introduced by opacity broadening of the lines at large
opacities.

\paragraph{Variation with Column Density.}

The line ratio behavior with increasing column density per unit
velocity dispersion is straightforward.  At the lowest column density,
all lines are optically thin and thermalized, or nearly so.  At a
temperature of 30~K, the $J = 3$ level is slightly more populated than
the $J = 2$ level, so the opacity and hence the line intensity is
higher for the $J$ = 3$\rightarrow$2 line.  As the column density
increases, the ($J$ = 3$\rightarrow$2)/($J$ = 2$\rightarrow$1) ratios
decrease as the lines become optically thick.  The $^{12}$CO/$^{13}$CO
ratios begin at a value close to their abundance ratio, since all the
lines are thin and thermalized at 30~K, and decrease to a ratio of
unity as the column density increases and the line intensities plateau
at their optically thick levels.

\paragraph{Variation with Density.}

At a density of 100 cm$^{-3}$ all lines are subthermally excited and
the population of the $J = 3$ level is at its lowest value relative to
the population of the $J = 2$ level.  Therefore the ($J$ =
3$\rightarrow$2)/($J$ = 2$\rightarrow$1) ratios increase as the density
increases since the excitation increases, and the ratios plateau at the
density at which both transitions are thermalized.  In contrast to the
($J$ = 3$\rightarrow$2)/($J$ = 2$\rightarrow$1) ratios, the
$^{12}$CO/$^{13}$CO ratios decrease with increasing density.  The
$^{12}$CO lines are highly optically thick, while the $^{13}$CO lines
have opacities of only a few or less.  Although all four line
intensities increase with density, the excitation temperature of the
$^{13}$CO lines rises faster than the $^{12}$CO lines (since radiative
trapping is less effective in the $^{13}$CO lines), causing the
$^{13}$CO line intensities to increase faster than the highly optically
thick $^{12}$CO lines.  As the level populations thermalize, the line
intensities remain constant and the line ratios plateau.

\paragraph{Variation with Temperature.}

At this density and column density, all four lines (CO $J$ =
2$\rightarrow$1, $^{13}$CO $J$ = 2$\rightarrow$1, CO $J$ =
3$\rightarrow$2, and $^{13}$CO $J$ = 3$\rightarrow$2) are thermalized.
The $^{12}$CO lines are highly optically thick at all temperatures,
while the $^{13}$CO lines are only marginally thick at the lowest
temperatures and become thin at the higher temperatures.  As the
temperature increases, the line opacities decrease as the upper $J$
levels become more populated, thus depopulating the $J = 2$ and $J = 3$
levels.  The ($J$ = 3$\rightarrow$2)/($J$ = 2$\rightarrow$1) ratios
increase slightly as $T$ increases because of the decreasing importance
of the Rayleigh-Jeans correction to the line radiation temperatures.
The $^{12}$CO/$^{13}$CO ratios increase since the $^{13}$CO lines
become more optically thin as $T$ increases but the $^{12}$CO lines
remain thick.  It is also apparent that line ratios involving the
isotopomer are more sensitive to changes in temperature.

\subsection{Physical Conditions}

The $^{12}$CO/$^{13}$CO $J$ = 2$\rightarrow$1 line ratio measured from
the observed peak temperatures is clearly correlated with the $^{13}$CO
peak temperature (Figure~\ref{fig-5});  a similar trend is seen in the
$J$ = 3$\rightarrow$2 transitions.  An increase in the peak temperature
line ratio can be produced by a lower column density, a higher kinetic
temperature, or a lower volume density.  In fact, the observed
correlation of the line ratio with peak temperature can be produced by
varying either the density or the column density while holding the
other variable constant. It is impossible to determine whether density
or column density variations are responsible for the observed trends
using the $J$ = 2$\rightarrow$1 data alone. However, the relatively
constant and large value of the $^{13}$CO $J$ =
(3$\rightarrow$2)/(2$\rightarrow$1) measured at six positions in the
cloud (Table~\ref{tbl-6}) suggests that the primary factor is
variations in the column density.  A comparison of the line ratios with
LVG models shows that values for the $^{13}$CO $J$ =
(3$\rightarrow$2)/(2$\rightarrow$1) line ratio greater than 1 can be
obtained for volume densities of the order of 10$^5$ cm$^{-3}$ and
kinetic temperatures in the range of 30--50 K or more. In contrast, to
explain the correlation in Figure~\ref{fig-5} using density variations
would require the density to vary from a low of $\sim 100$ cm$^{-3}$ to
at least $10^4$ cm$^{-3}$. Such a low density is incompatible with the
high $^{13}$CO $J$ = (3$\rightarrow$2)/(2$\rightarrow$1) line ratio
observed, especially for the weakest position in the cloud (at offset
(2,12), Table~\ref{tbl-6}).  Clearly, more observations of the
$^{13}$CO $J$ = (3$\rightarrow$2)/(2$\rightarrow$1) line ratio in
regions of M17 with low peak temperatures would be helpful to rule out
the possibility of density variations more conclusively. However, note
that similarly large $^{13}$CO $J$ =
(3$\rightarrow$2)/(2$\rightarrow$1) line ratios have also been observed
in the Orion B and Rosette molecular clouds (\markcite{k96}Kramer et
al.\ 1996; \markcite{s98}Schneider et al.\ 1998).

The spread in the data can be explained by variations in the kinetic
temperature from low values of 10--20 K (where the lower limit is not
well-constrained by these data) up to about 50 K (Figure~\ref{fig-5}).
Thus, the simplest interpretation of the trends shown in
Figure~\ref{fig-5} is that most of the cloud is at relatively large
densities, with variations in the $^{12}$CO column densities per unit
velocity from $10^{17}$ to $10^{18}$ cm$^{-2}$ (km s$^{-1}$)$^{-1}$ and
in the kinetic temperature from 10--20 to about 50 K explaining the
broad correlation and the scatter about this correlation,
respectively.  This picture is quite different from the
\markcite{s94}Sakamoto et al.\ (1994) interpretation of varying density
being responsible for the $^{12}$CO ($J$ = 2$\rightarrow$1)/($J$ =
1$\rightarrow$0) line ratio variations in the Orion molecular cloud.
One possible explanation is that our M17 map may be dominated from
high-density emission in the center of the molecular cloud, and may not
extend far enough into the edges of the cloud 
to pick up the effect of lower-density
halo material on the $^{12}$CO line ratio.

We used the LVG code RAD written by Lee Mundy and implemented as part
of the MIRIAD data reduction package (\markcite{stw95}Sault, Teuben, \&
Wright 1995) to estimate the physical conditions in M17.  LVG models
(e.g., \markcite{ss74}Scoville \& Solomon 1974), microturbulent models
(\markcite{w77}White 1977), and photon mean escape probability
formalisms often yield similar results when applied to galactic
molecular clouds (\markcite{w77}White 1977).  We incorporated the
collision cross-section rates of \markcite{fl85}Flower \& Launay (1985)
and \markcite{sch85}Schinke et al.\ (1985) (para-H$_2$ with CO) into
the LVG models.  Models were run for kinetic temperatures $T_K =
10-1000$ K, in logarithmic intervals of 0.1.  The model density range
is $n_{H_2} = 10^2-10^6$ cm$^{-3}$ and the $^{12}$CO column density
range per unit velocity is $N(^{12}\rm CO)/\Delta V = 10^{16}-10^{21}$
cm$^{-2}$ (km s$^{-1}$)$^{-1}$, both in logarithmic intervals of 0.25.
To calculate the $^{12}$CO column density from $N/\Delta V$, we
approximate $\Delta V$ by the ratio of the $^{13}$CO $J$ =
2$\rightarrow$1 intensity to the peak $^{13}$CO temperature, $\Delta V
= I(^{13}{\rm CO})/T_{peak}$. We adopt a [$^{12}$CO]/[$^{13}$CO]
abundance ratio of 50 for M17, which is similar to the value measured
in W51 which is at a similar Galactic radius (\markcite{lp90}Langer \&
Penzias 1990).  Model line ratios and $\chi^2$ values at each model
grid point for each position were generated using the MIRIAD ``maths''
command.

Using the LVG models, we determined the physical conditions for six of
the nine positions for which observations were made with the same beam
in all four lines. The observed $^{12}$CO line widths are significantly
broader than the $^{13}$CO line widths, probably due to opacity
broadening of the $^{12}$CO lines. For this reason, we used the
observed peak temperatures at each position to determine the line
ratios rather than the integrated intensities.  Three of the nine
positions show evidence for self-absorption in the $^{12}$CO $J$ =
3$\rightarrow$2 and $J$ = 2$\rightarrow$1 lines, with the $^{13}$CO
peak lying between two $^{12}$CO peaks. This self-absorption makes it
impossible to model these three positions using the simple
single-component LVG formalism.  We also used our best estimate of the
global CO line ratios in M17 to see how well the results from a global
average compare with the results obtained at individual locations.  The
$^{12}$CO $J$ = 2$\rightarrow$1 line averaged over all the observed
spectra is only 10\% wider than the $^{13}$CO $J$ = 2$\rightarrow$1
line measured in the same manner, and so it is reasonable to use the
global line ratios obtained from integrated intensities in the LVG
analysis. These similar linewidths show that, when averaged over
the entire cloud, the observed broadening of these two lines is dominated
by the kinematics of the gas, rather than the opacity.

Best-fit solutions for each set of four line ratios were found by
searching the $\chi^2$ model cubes for their minimum value.  Acceptable
solutions were defined to be ones for which the minimum $\chi^2$ value
was less than 4.0, and for which the $^{13}$CO $J$ = 2$\rightarrow$1
peak temperature was less than or equal to the value predicted by the
model. (This last constraint was not used in fitting the global line
ratios.) As an example, the solution for the line ratios averaged over
the entire cloud is shown in Figure~\ref{fig-6}.  The best-fit
solutions are given in Table~\ref{tbl-6}. In general, the best fitting
combinations of temperature, density, and column density are correlated
in the sense that solutions with higher temperatures have higher column
densities and lower densities.  For positions (2\arcmin,6\arcmin),
(2\arcmin,$-$12\arcmin), and ($-$4\arcmin,6\arcmin), the minimum
$\chi^2$ values occurred at temperatures of about 1000~K, but these
solutions resulted in a combination of density and column density that
gave unreasonably long columns of gas and required the M17 cloud to be
highly elongated along the line of sight.  For this reason, we chose
$\chi^2$ minima restricted to models with $T_K \leq 250$~K.

Acceptable solutions are found for five of the six individual locations
fitted with the LVG models.  The one location at (2\arcmin,6\arcmin)
for which no solution is found has a rather low $^{12}$CO ($J$ =
3$\rightarrow$2)/($J$ = 2$\rightarrow$1) line ratio which cannot be fit
in tandem with the other three ratios.  For all of these locations, a
kinetic temperature in the range of 25 to 50 K and higher produces an
acceptable solution.  At the best-fit temperature for each position,
any density greater than about $2\times 10^4$ cm$^{-3}$ produces an
acceptable fit. The $^{12}$CO column density is typically in the range
of $1\times 10^{18}$ to $2\times 10^{18}$ cm$^{-2}$. Interestingly, the
observed $^{13}$CO $J$ = 2$\rightarrow$1 peak brightness temperatures
agree very well with the model temperatures for four of the five
locations. This result suggests that the emitting gas fills the
22\arcsec\ (0.2 pc) beam of the JCMT.

An acceptable solution with a kinetic temperature of 30~K is also
found using the estimates of the global line ratios (Table~\ref{tbl-6}
and Figure~\ref{fig-6}). The results from the estimates of the global
line ratios are in reasonable agreement with the estimates obtained at
the individual locations around the cloud. This agreement suggests that
observations of CO line ratios integrated over entire molecular clouds
in nearby galaxies are likely to give physically meaningful
measurements of the average physical conditions in the clouds.

Most of the previous studies of the density, temperature, and column
density of M17 have focused on the region around the PDR, with
temperature determinations in the range of 45 to $\ge 200$ K
(\markcite{s88}Stutzki et al.\ 1988; \markcite{h87}Harris et al.\ 1987;
\markcite{g92}Greaves et al.\ 1992) and densities in the range of
$10^5$ to $10^6$ cm$^{-3}$ (\markcite{m86}Mundy et al.\ 1986;
\markcite{w93}Wang et al.\ 1993; \markcite{sg90}Stutzki \& G\"usten
1990).  Only two studies address the question of the physical
conditions in M17 outside the PDR region. \markcite{s88}Stutzki et al.\ (1988)
find that most of the column density of the cloud is at a kinetic
temperature of 50 K, while \markcite{g92}Greaves et al.\ (1992) give a
kinetic temperature of 30--45 K away from the emission peak of the
PDR.  These temperatures are in general agreement with the temperature
of 30~K obtained from our LVG solution using the global averaged CO
line ratios.  Higher temperatures obtained in the previous studies are
likely due to their relatively limited spatial sampling (a
4--6\arcmin\ strip map through the PDR peak, and a $2\times 2^\prime$
region centered on the PDR, respectively) and the fact that the region
studied was very near to the PDR, which, as we have seen, has higher
temperatures than most regions of the cloud. 

The density limits obtained here for M17 agree quite well with the
densities of $\sim 10^5$ cm$^{-3}$ required to explain the observed $J$
= 2$\rightarrow$1 and $J$ = 3$\rightarrow$2 line ratios in the Orion B
molecular cloud (\markcite{k96}Kramer et al.\ 1996), although the
densities derived for the Rosette molecular cloud are typically
somewhat lower ($\sim 10^3-10^4$, \markcite{s98}Schneider et
al.\ 1998).  The average kinetic temperature and $^{12}$CO column
density for M17 obtained here are similar to the temperatures and
column densities obtained for individual molecular clouds in the Local
Group galaxy M33 (\markcite{w97}Wilson et al.\ 1997). However, the
density solutions for M17 are generally larger by at least an order of
magnitude than the solutions for the M33 clouds.  It is interesting
that the solutions for the molecular cloud in IC 10 for which $^{13}$CO
$J$ = 3$\rightarrow$2 data are available give a density very similar to
the lower limits derived here for M17 ($10^4-10^5$ cm$^{-3}$,
\markcite{pw98}Petitpas \& Wilson 1998).  However, that cloud in IC 10
also has a higher kinetic temperature (100 K) and column density
($6\times 10^{18}$ cm$^{-2}$) than the average M17 solution.

\section{Implications for Determining Physical Properties in Other Galaxies}

\subsection{The Dependence of Cloud Size on the Emission Line}

One of the important issues in attempting to derive physical conditions
in external galaxies is whether or not the beam filling factors of the
different CO emission lines are the same. For example, if the $J$ =
3$\rightarrow$2 transitions are confined to a smaller region of a cloud
than the $J$ = 2$\rightarrow$1 transitions, observations that do not
resolve the cloud will underestimate the true ($J$ =
3$\rightarrow$2)/($J$ = 2$\rightarrow$1) line ratio for the $J$ =
3$\rightarrow$2 emitting regions.  On the other hand, we might argue
that global measurements, such as those presented here for M17, provide
the best estimate of the {\it average } conditions in molecular clouds;
such global measurements are not affected by the beam filling problem.
In essence, most extragalactic work on line ratios implicitly assumes
constant line ratios across the cloud, while in extreme cases (such as
the comparison of the $J$ = 6$\rightarrow$5 and $J$ = 1$\rightarrow$0
transitions), one of the lines may be confined to only a very small
area of the cloud.  The large-area maps of M17 presented in this paper
provide us with the first opportunity to assess how significantly a
real cloud deviates from the idealized picture of spatially constant
line ratios.

To attempt a quantitative comparison, we note that one estimate of the
relative size of a source in different emission lines can be obtained
by measuring the ratio of the observed peak brightness temperature
divided by the average brightness temperature over the same area for
each emission line, and comparing these temperature ratios for the
different emission lines.  (In this method, an emission line which is
less extended on the sky has a lower measured average brightness
temperature, because many regions in the averaging area contain no
emission.) Comparing our three maps, we obtain $\sim 1.5$ for the
$^{12}$CO/$^{13}$CO $J$ = 2$\rightarrow$1 ``size'' ratio, consistent
with the trend of increasing $^{12}$CO/$^{13}$CO line ratio that was
seen in section \S3.2.  The $^{13}$CO ($J$ = 1$\rightarrow$0)/($J$ =
2$\rightarrow$1) ``size'' ratio is only $\sim 1.2$, however, probably
within the noise of most extragalactic line ratio determinations.
Thus, for the two lowest rotational transitions, the filling factor
correction is small for ratios of the line transitions, but may be
significant for the isotopic line ratios.

The situation is more difficult to assess for the $J$ = 3$\rightarrow$2
transitions, since we did not make complete maps of M17 in these
transitions. The nine positions for which the $^{12}$CO ($J$ =
3$\rightarrow$2)/($J$ = 2$\rightarrow$1) line ratio have been measured
have very similar line ratios, including two positions at (2\arcmin,
12\arcmin) and (0\arcmin,12\arcmin) that lie below the 10\% of peak
contour in the $^{13}$CO $J$ = 1$\rightarrow$0 map. Thus, it seems likely
that M17 has a similar spatial extent in the $J$ = 3$\rightarrow$2 and
$J$ = 2$\rightarrow$1 transitions of $^{12}$CO, but it would be useful
to confirm this result with large-area $J$ = 3$\rightarrow$2 maps.  The
cuts through the cloud in the $J$ = 3$\rightarrow$2 transitions
(Figure~\ref{fig-2}) show that the $^{12}$CO/$^{13}$CO $J$ =
3$\rightarrow$2 line ratio is usually larger in the outer regions of
the cloud, while the line ratio shows a similar trend with integrated
intensity or peak temperature as does the $J$ = 2$\rightarrow$1 transition
(cf. Figure~\ref{fig-5}).  These observations strongly suggest that the
$^{13}$CO $J$ = 3$\rightarrow$2 transition is more compact in its
spatial extent than the $^{12}$CO $J$ = 3$\rightarrow$2 transition,
although, again, it would be useful to confirm this result with a
larger area map.

It is clear from these data that the size of a molecular cloud, and
hence its filling factor in a large beam, depends on the transition and
isotopomer being used to study the cloud. However, the LVG analysis
discussed above shows that physical conditions derived from
globally-averaged line ratios give good agreement with the physical
conditions determined along individual lines of sight through the
cloud. This agreement suggests that the different filling factors of
the low-J CO transitions will not produce large systematic errors
in the derived physical conditions in normal extragalactic systems.

\subsection{Evidence for a Significant Low Column Density Component in the
Molecular Interstellar Medium}

How do the global CO line ratios we have obtained for M17 compare with
measurements of these same line ratios in external galaxies?  The best
agreement is for the $^{12}$CO ($J$ = 3$\rightarrow$2)/($J$ =
2$\rightarrow$1) line ratio.  The M17 line ratio of $0.76\pm0.19$
agrees quite well with the values obtained in normal molecular clouds
in M33 ($0.69\pm0.06$, \markcite{w97}Wilson et al.\ 1997), in the
starburst galaxies M82 ($0.8\pm0.2$, \markcite{g93}G\"usten et
al.\ 1993), NGC 253 ($0.5\pm0.1$ in the disk, \markcite{w91}Wall et
al.\ 1991), and IC 342 ($0.47\pm0.12$, \markcite{ia93}Irwin \& Avery
1993), and in the normal galaxy M51 (0.7, \markcite{gb93}Garcia-Burillo
et al.\ 1993).  The relatively small range of values observed for
the $^{12}$CO ($J$ = 3$\rightarrow$2)/($J$ =
2$\rightarrow$1) line ratio is not unexpected, given the weak
dependence of this line ratio on physical conditions (Figure~\ref{fig-4}).
Due to the very weak line strengths, there are few
extragalactic observations of the $^{13}$CO ($J$ =
3$\rightarrow$2)/($J$ = 2$\rightarrow$1) line ratio.  One measurement
in NGC 253 gives a line ratio of $2.0\pm0.6$ in the starburst nucleus
and $0.86\pm0.14$ for the average of three positions offset from the
nucleus by 10\arcsec\ or $\sim 120$ pc (\markcite{w91}Wall et
al.\ 1991).  Given the large measurement uncertainties in these weak
extragalactic lines, the agreement with the M17 line ratio of
$1.3\pm0.3$ is quite good.

The agreement of the M17 $^{12}$CO/$^{13}$CO line ratios with
extragalactic measurements is not as good. There are now measurements
of the $^{12}$CO/$^{13}$CO $J$ = 2$\rightarrow$1 line ratio for many
galaxies. These line ratios range from a low of $7.3\pm 0.5$ in M33
(\markcite{w97}Wilson et al.\ 1997) to a high of $13\pm 5$ for a large
sample of starburst and interacting galaxies (\markcite{a95}Aalto et
al.\ 1995).  These values are all significantly larger than the value
of $4.5\pm0.7$ that is found here for M17.  Thus, the
$^{12}$CO/$^{13}$CO $J$ = 2$\rightarrow$1 ratio appears to increase as
we move from individual Galactic molecular clouds to normal galaxies to
starburst galaxies. A similar trend is also seen for the
$^{12}$CO/$^{13}$CO $J$ = 1$\rightarrow$0 line ratio, where the line
ratio measured in Galactic molecular clouds is 3--6
(\markcite{gb76}Gordon \& Burton 1976; \markcite{sss79}Solomon,
Scoville, \& Sanders 1979), the ratio in normal galaxies is $\sim 10$
(\markcite{si91}Sage \& Isbell 1991), and the ratio is even larger in
starburst galaxies (i.e., \markcite{a91}Aalto et al.\ 1991).  There are
only a few extragalactic measurements of the $^{12}$CO/$^{13}$CO $J$ =
3$\rightarrow$2 line ratio, but they appear to follow the same general
trend seen for the $J$ = 2$\rightarrow$1 and $J$ = 1$\rightarrow$0 line
ratios.  For example, the $^{12}$CO/$^{13}$CO $J$ = 3$\rightarrow$2
line ratio measured for a single molecular cloud in the Local Group
irregular galaxy IC 10 is $5.3 \pm 0.4$ (\markcite{pw98}Petitpas \&
Wilson 1998), while the average value for five positions in the nucleus
of the starburst galaxy NGC 253 is $17\pm 6$ (\markcite{w91}Wall et
al.\ 1991).  Both these values are larger than the global value of
$3.7\pm 0.9$ determined here for M17.

It may be possible to explain the larger line ratios in the starburst
systems by a difference in the [$^{12}$CO]/[$^{13}$CO] abundance ratio
(e.g., \markcite{c92}Casoli, Dupraz, \& Combes 1992) or by the unusual
physical conditions in these hot, dense regions (\markcite{a97}Aalto et
al.\ 1997).  However, for normal galaxies, the higher
$^{12}$CO/$^{13}$CO line ratios are most likely caused by the emission
in the beam originating from a combination of emission from both giant
molecular clouds and lower column density material (\markcite{p88}Polk
et al.\ 1988; \markcite{ww94}Wilson \& Walker 1994). This low column
density material might be the envelopes of giant molecular clouds, or
separate low-density clouds such as translucent or high latitude
clouds.  In this scenario, the higher average line ratio arises because
the low column density material has a larger intrinsic line ratio,
which can approach the true abundance ratio in the optically thin
limit. An extended low column density component has been identified in
the Milky Way from $J$ = 1$\rightarrow$0 observations of the Gemini OB1
molecular complex (\markcite{c95}Carpenter, Snell, \& Schoerb 1995).
These data show that 50\% of the total CO luminosity of Gem OB1, but
only 20\% of its mass, is contained in this extended component, which
typically has H$_2$ column densities less than $2\times 10^{21}$
cm$^{-2}$. If the Gem OB1 complex is typical of molecular clouds
elsewhere in the galaxy, this low column density gas may contribute a
significant fraction of the $^{12}$CO luminosity of the interstellar
medium (\markcite{c95}Carpenter et al.\ 1995). The observations of M17
presented here, when combined with extragalactic observations of the
$^{12}$CO/$^{13}$CO $J$ = 2$\rightarrow$1 and $J$ = 3$\rightarrow$2
line ratios, strongly suggest that molecular gas with low column
densities contributes significant luminosity in the $J$ =
2$\rightarrow$1 and $J$ = 3$\rightarrow$2 transitions as well.

The possible presence of a significant low column density component to
the molecular interstellar medium introduces a significant source of
uncertainty into studies of other galaxies, since the observed CO line
strengths will depend on the relative contribution of low and high
column density material within the beam.  Observational support for
this difficulty is provided by high-resolution observations of M51,
which show that the $^{12}$CO/$^{13}$CO $J$ = 2$\rightarrow$1 ratio is
significantly larger in the interarm regions than in the arm regions
(\markcite{gb93}Garcia-Burillo et al.\ 1993). Due to the impossibility
of fitting the on-arm line ratios with a single component LVG model,
the authors interpret these line ratios as evidence for primarily cold,
low-density clouds in the interarm region, but require a mixture of
these cloud ``halos'' and denser warmer ``cores'' in the arms to fit
the observed line ratios.  If we wish to study variations in the
physical properties of the dense, star-forming molecular clouds,
interferometric measurements in the $J$ = 2$\rightarrow$1 and $J$ =
3$\rightarrow$2 transitions may be required to isolate the emission
from giant molecular clouds from surrounding diffuse molecular
material.  It will also be important to determine whether starburst
galaxies can also be understood in this general framework of high line
ratios indicating extensive low column density material, albeit perhaps
at a higher temperature due to the higher ultraviolet flux, or whether
some other physical mechanism is necessary in these unusual
environments.

\section{Conclusions}

In this paper we have presented large-area maps of the Galactic
molecular cloud M17 in the $^{12}$CO and $^{13}$CO $J$ =
3$\rightarrow$2 and $J$ = 2$\rightarrow$1 transitions. The $J$ =
2$\rightarrow$1 data cover a region of $16\arcmin \times 32\arcmin$
($10 \times 20$ pc) with 2\arcmin\ sampling, while the $J$ =
3$\rightarrow$2 data consist of four cuts through the cloud as well as
spectra at four additional locations in the cloud.  We have used these
data sets to study the large-scale variation in the CO line ratios
across the cloud, as well as to obtain a good estimate of the globally
averaged line ratios and physical conditions of the cloud.  Our main
results are summarized below.

(1) Both the $^{12}$CO/$^{13}$CO $J$ = 2$\rightarrow$1 and the
$^{12}$CO/$^{13}$CO $J$ = 3$\rightarrow$2 line ratios show a clear
correlation with the $^{13}$CO integrated intensity, in the sense that
the line ratio is smaller for locations with large integrated
intensities. This correlation has its origins in the varying column
density of the gas from one position to another, with some variation in
the kinetic temperature also
playing a role.

(2) In contrast to the $^{12}$CO/$^{13}$CO line ratios, the $^{12}$CO
($J$ = 3$\rightarrow$2)/($J$ = 2$\rightarrow$1) and the $^{13}$CO ($J$
= 3$\rightarrow$2)/($J$ = 2$\rightarrow$1) line ratios show no
significant variation from place to place within M17. In fact, no
significant change in these line ratios is seen even on the peak of the
PDR, where published data suggest some gas with a temperature of at
least 200 K is present (\markcite{h87}Harris et al.\ 1987).

(3) We have used the data to obtain estimates for the various line
ratios averaged over the entire cloud. Our best estimates for the
global line ratios are as follows:  $^{12}$CO/$^{13}$CO $J$ =
2$\rightarrow$1: $4.5\pm0.7$; $^{12}$CO/$^{13}$CO $J$ =
3$\rightarrow$2: $3.7\pm0.9$; $^{12}$CO ($J$ = 3$\rightarrow$2)/($J$ =
2$\rightarrow$1): $0.76\pm0.19$; $^{13}$CO ($J$ = 3$\rightarrow$2)/($J$
= 2$\rightarrow$1): $1.3\pm0.3$. These line ratios generally agree
quite well with previous measurements over smaller regions of Galactic
molecular clouds and with results obtained from a pencil-beam survey of
the Galactic plane (\markcite{s93}Sanders et al.\ 1993), with the
exception of the $^{12}$CO/$^{13}$CO $J$ = 3$\rightarrow$2 and
$^{13}$CO ($J$ = 3$\rightarrow$2)/($J$ = 2$\rightarrow$1) line ratios,
which are substantially larger and smaller, respectively, in the
Galactic plane survey than in M17.

(4) We have combined beam-matched observations of all four transitions
with an LVG code to determine the temperature, density, and column
density at five locations around the cloud.  The minimum $\chi^2$
solution using the globally averaged line ratios gives a kinetic
temperature of 30 K, a density of $0.2-5\times 10^5$ cm$^{-3}$, and a
$^{12}$CO column density of $1\times 10^{18}$ cm$^{-2}$. This result is
in reasonable agreement with the results obtained for the individual
lines-of-sight through the cloud, which suggests that the typical
physical conditions in a molecular cloud can be determined using CO
line ratios integrated over the entire cloud.  Thus, observations of CO
line ratios in non-starburst external galaxies which integrate the
emission over an entire molecular cloud, or perhaps even ensembles of
molecular clouds, are likely to give physically meaningful results.

(5) A comparison of the M17 results with published observations for
external galaxies reveals that the $^{12}$CO/$^{13}$CO $J$ =
2$\rightarrow$1 and $J$ = 3$\rightarrow$2 line ratios are
systematically larger in external galaxies than in M17. There appears
to be a clear trend of increasing line ratios as one moves from
Galactic molecular cloud cores to entire Galactic molecular clouds to
normal galaxies to starburst galaxies. A similar trend has been noted
previously for the $^{12}$CO/$^{13}$CO $J$ = 1$\rightarrow$0 line ratio
(\markcite{si91}Sage \& Isbell 1991; \markcite{a91}Aalto et al.\ 1991).
The high line ratios in starburst galaxies may be due to the unusual
conditions in these dense, hot regions (\markcite{a97}Aalto et
al.\ 1997), while for normal galaxies the most likely explanation is a
significant contribution to the CO emission by low column density material
(\markcite{ww94}Wilson \& Walker 1994). These new observations of M17
show that the difference between the $^{12}$CO/$^{13}$CO line ratios
for Galactic molecular clouds and the disks of spiral galaxies occurs
for all three of the lowest rotational transitions. These results show
that care must be taken in interpreting line ratio variations from
place to place within a galaxy, as they could be due to changes in the
relative contribution of low and high column density material, rather
than to changes in the temperature or other physical conditions in the
dense molecular gas.

\acknowledgments
The research of CDW is supported through a grant from the Natural
Sciences and Engineering Research Council of Canada. We thank Jennifer
Larking for assistance with the data reduction. The JCMT is operated by
the Royal Observatories on behalf of the Particle Physics and Astronomy
Research Council of the United Kingdom, the Netherlands Organization
for Scientific Research, and the National Research Council of Canada.
The Five College Radio Astronomy Observatory is operated with the
permission of the Metropolitan District Commission, Commonwealth of
Massachusetts, and with the support of the National Science Foundation
under grant AST-9420159.

\appendix

\section{Calculating Global Line Ratios in M17}

Since we have a uniformly sampled map of the M17 molecular cloud, it is
relatively straightforward to obtain an estimate of the global
$^{12}$CO/$^{13}$CO $J$ = 2$\rightarrow$1 line ratio, i.e., the line
ratio averaged over the entire cloud.
A good estimate of the global line ratio can be obtained by simply
summing the $^{12}$CO $J$ = 2$\rightarrow$1 integrated intensity at each
position, performing a similar summation for the $^{13}$CO line, and
taking the ratio of the two summed intensities. This method gives a
global $^{12}$CO/$^{13}$CO $J$ = 2$\rightarrow$1 line ratio of $4.5\pm
0.7$, where the uncertainty is the 15\% uncertainty in the absolute
calibration (\S 2).

We use the beam-matched $J$ = 2$\rightarrow$1 and $J$ =
3$\rightarrow$2 observations for nine positions in M17 to determine the
$^{12}$CO and $^{13}$CO ($J$ = 3$\rightarrow$2)/($J$ = 2$\rightarrow$1)
line ratios (Table~\ref{tbl-4}). Five of the positions are located
along the Declination cut with 6\arcmin\ spacing passing through the PDR,
while the remaining four positions are located at the peak emission of
M17 North, near the western edge of each of M17 South and M17 North
(away from the PDR), and near the northern edge of M17 North.  The
$^{12}$CO ($J$ = 3$\rightarrow$2)/($J$ = 2$\rightarrow$1) line ratio
appears very constant across all nine positions, with an rms dispersion
of only 10\%. Including the absolute calibration uncertainties, the
average value obtained from summing the integrated intensities in the
two transitions is $0.76\pm0.19$, which we adopt also as the best
measurement of the global average value of this line ratio in M17.  The
$^{13}$CO ($J$ = 3$\rightarrow$2)/($J$ = 2$\rightarrow$1) line ratio
also appears fairly constant across the nine positions, with an rms
dispersion of 17\%, at least a factor of two smaller than the
dispersions in the $^{12}$CO/$^{13}$CO line ratios.
The average of the individual line ratio determinations is 1.2, while
the ratio of the sums of the $J$ = 3$\rightarrow$2 and $J$ =
2$\rightarrow$1 integrated intensities is 1.3. The good agreement
between these two methods suggests that the large contribution to the
integrated intensity from the PDR and M17 North is not biasing the
results too badly compared to what would be obtained from a better-sampled map,
and so we adopt $1.3\pm0.3$ as our estimate of the global $^{13}$CO
($J$ = 3$\rightarrow$2)/($J$ = 2$\rightarrow$1) line ratio in M17.

Unlike the situation for the $J$ = 2$\rightarrow$1 data, a simple sum
of the $^{12}$CO and $^{13}$CO integrated intensities for all the
spectra will not produce a good estimate of the global
$^{12}$CO/$^{13}$CO $J$ = 3$\rightarrow$2 line ratio in M17, since the
strong emission from the PDR would be given too much weight. Instead,
we use a simple average of the line ratios at the 23 positions with
good detections of the $^{13}$CO line, which gives a value of
$3.7\pm0.9$. We adopt a somewhat higher uncertainty of 25\% to reflect
the highly undersampled nature of the $J$ = 3$\rightarrow$2 data. As a
check on the validity of this simple average, we also estimated the
global $^{12}$CO/$^{13}$CO $J$ = 3$\rightarrow$2 line ratio in M17 by
using the good correlation seen between the $^{13}$CO $J$ =
3$\rightarrow$2 integrated intensity and the $^{12}$CO/$^{13}$CO $J$ =
3$\rightarrow$2 line ratio (Figure~\ref{fig-5}, 
as well as the fact that the $^{12}$CO ($J$
= 3$\rightarrow$2)/($J$ = 2$\rightarrow$1) line ratio appears to be
constant throughout the cloud. We first calculated the $^{12}$CO $J$ =
3$\rightarrow$2 integrated intensity by scaling the $^{12}$CO $J$ =
2$\rightarrow$1 integrated intensity map by the average $^{12}$CO ($J$
= 3$\rightarrow$2)/($J$ = 2$\rightarrow$1) line ratio (0.76). We then
calculated the $^{13}$CO $J$ = 3$\rightarrow$2 integrated intensity at
each position using the relation $I(^{13}\rm{CO}) =
(I(^{12}\rm{CO})/9.8)^{1.5}$ derived from a least squares fit to the
data points.  Finally, the global $^{12}$CO/$^{13}$CO $J$ =
3$\rightarrow$2 line ratio was calculated from the ratio of the sums of
the $^{12}$CO and $^{13}$CO integrated intensities.  This procedure
yielded a best estimate of the global $^{12}$CO/$^{13}$CO $J$ =
3$\rightarrow$2 line ratio in M17 of $3.2\pm0.8$, which suggests that
the value derived from averaging the individual line ratios is not
biased too heavily by the undersampled nature of the map.

\clearpage

\figcaption[m17co_fig1.ps]{The $^{12}$CO $J$ = 2$\rightarrow$1,
$^{13}$CO $J$ = 2$\rightarrow$1, and $^{13}$CO $J$ = 1$\rightarrow$0 integrated
intensity of the M17 giant molecular cloud, measured in the hybrid
temperature scale described in \S 2 over the velocity range 10 to 30 km
s$^{-1}$.  The (0,0) position corresponds to ($\alpha(1950)$ = 18$^{\rm
h}$17$^{\rm m} $21\fs0, $\delta(1950)$ = $-$16\arcdeg 06\arcmin
45\arcsec) and the peak of the photon-dominated region M17 SW lies at
offset (2,$-$6).  The contour levels are 5, 10, 15, 20, 30, 40, 50, 60,
70, 80, and 90 percent of the map peak intensity.  (a) The
$^{12}$CO $J$ = 2$\rightarrow$1 integrated intensity map. Spectra were
obtained in a 22\arcsec\ beam and sampled on a 2\arcmin\ grid, shown as
dots in the figure. The map was produced by interpolating the original
data onto a 30\arcsec\ grid using a bicubic spline algorithm.  The map
peak intensity is 639 K km s$^{-1}$, about 7 percent higher than the
peak intensity measured in the 22\arcsec\ JCMT beam.  (b) The
$^{13}$CO $J$ = 2$\rightarrow$1 integrated intensity map. The map was
produced in the same manner as the $^{12}$CO map.  The map peak
intensity is 233 K km s$^{-1}$, about 10 percent higher than the peak
intensity measured in the 22\arcsec\ JCMT beam.  (c) The
$^{13}$CO $J$ = 1$\rightarrow$0 integrated intensity map.  Spectra were
obtained in a 47\arcsec\ beam sampled on a 25\arcsec\ grid.  The data
were smoothed to 1\arcmin\ resolution in producing the final map.  The
map peak intensity is 126 K km s$^{-1}$.  Note the good correspondence
in the large-scale structures between the undersampled $J$ =
2$\rightarrow$1 maps and this nearly fully sampled map.
\label{fig-1}}

\figcaption[m17co_fig2.ps]{The $^{12}$CO/$^{13}$CO $J$ =
3$\rightarrow$2 integrated intensity 
ratio for cuts at three different declinations through
M17. The ratio was measured in a 15\arcsec\ beam.  The solid line
connects measurements made at 0\arcmin\ declination offset, the dashed
line connects measurements made at $-$8\arcmin\ declination offset, and
the dotted line connects measurements made at 8\arcmin\ declination
offset. The error bars indicate a 15\% calibration uncertainty.  The
(0,0) position corresponds to ($\alpha(1950)$ = 18$^{\rm h}$17$^{\rm
m}$21\fs0, $\delta(1950)$ = $-$16\arcdeg 06\arcmin 45\arcsec).
\label{fig-2}}

\figcaption[m17co_fig3.ps]{Examples of $^{12}$CO and $^{13}$CO $J$ =
3$\rightarrow$2 spectra. The dashed line indicates the $^{12}$CO
spectrum and the solid line indicates the $^{13}$CO spectrum.  (a) A
typical pair of spectra located at (4\arcmin,0\arcmin) showing similar
$^{12}$CO/$^{13}$CO $J$ = 3$\rightarrow$2 line ratios in the two
velocity components. (b) Spectra at (2\arcmin,0\arcmin) showing
possible $^{12}$CO self absorption.  (c) Spectra at
(2\arcmin,$-$12\arcmin) near the south edge of the cloud showing
different $^{12}$CO/$^{13}$CO $J$ = 3$\rightarrow$2 line ratios in the
different velocity components, as well as $^{12}$CO self-absorption in
the higher velocity component.  (d) Spectra at ($-$2\arcmin,8\arcmin)
in M17 North showing strong $^{12}$CO but very weak $^{13}$CO.  (e)
Spectra of the photon-dominated region M17 SW showing very complex
emission structure, including possible $^{12}$CO self absorption at
$\sim 21$ km s$^{-1}$. Note the expanded vertical scale for this
panel.  \label{fig-3}}

\figcaption[m17co_fig4.ps]{Plots of LVG model results of the variation
of ratios of peak line temperature $T_R$.  For all plots the left
ordinate gives the scale for the $^{12}$CO and $^{13}$CO ($J$ =
3$\rightarrow$2)/($J$ = 2$\rightarrow$1) ratios (solid lines), and the
right ordinate gives the scale for the $^{12}$CO/$^{13}$CO $J$ =
2$\rightarrow$1 and $J$ = 3$\rightarrow$2 ratios (dashed lines).  (Top)
Variation of line ratios with the column density per unit velocity
dispersion, holding the kinetic temperature fixed at 30~K and the H$_2$
density fixed at $10^{5}$ cm$^{-3}$.  (Middle) Variation of line ratios
with H$_2$ density, holding the kinetic temperature fixed at 30~K and
the column density per unit velocity dispersion fixed at
$5\times10^{17}$ cm$^{-2}$ km$^{-1}$ s.  (Bottom) Variation of line
ratios with kinetic temperature, holding the column density per unit
velocity dispersion fixed at $5\times10^{17}$ cm$^{-2}$ km$^{-1}$ s and
the H$_2$ density fixed at $10^{5}$ cm$^{-3}$.
\label{fig-4}}

\figcaption[m17co_fig5.ps]{The $^{12}$CO$/^{13}$CO $J$ =
2$\rightarrow$1 peak temperature ratio as a function of the $^{13}$CO
$J$ = 2$\rightarrow$1 peak temperature (open squares), plotted on a
logarithmic scale. Both values were measured in a 22\arcsec\ beam.
Only spectra that were not affected by self-absorption or complex line
profiles are included.  The $^{12}$CO/$^{13}$CO $J$ = 3$\rightarrow$2
ratio as a function of the $^{13}$CO $J$ = 3$\rightarrow$2 peak
temperature, measured in a 15\arcsec\ beam, is also shown (filled
triangles). Overlaid on the figure are the predicted $J$ = 2$\rightarrow$1
relations from LVG models for three different values of the kinetic
temperature. The models assume a constant density of $10^5$ cm$^{-3}$.
The column density increases along each line towards the lower right
and is marked at logarithmic intervals of 0.33.
\label{fig-5}}

\figcaption[m17co_fig6.ps]{The LVG solution at a kinetic temperature
$T_K=30$~K for the CO line ratios for M17 averaged over the entire
cloud.  The lines indicate the observed $\pm 1\sigma$ values for the
line ratios of:  $^{12}$CO/$^{13}$CO $J$ = 2$\rightarrow$1 ($4.5\pm
0.7$, solid line); $^{12}$CO/$^{13}$CO $J$ = 3$\rightarrow$2 ($3.7\pm
0.9$, dashed line); $^{12}$CO ($J$ = 3$\rightarrow$2)/($J$ =
2$\rightarrow$1) ($0.76\pm 0.19$, dot-dashed line; lower limit in lower
left corner, upper limit along lower right); $^{13}$CO ($J$ =
3$\rightarrow$2)/($J$ = 2$\rightarrow$1) ($1.3\pm 0.3$, dotted line;
only the lower limit appears in the figure).  
The hatched area
indicates the region where all four sets of lines intersect, and the
black area encompasses the region where $\Delta\chi^2 \le 1$.
\label{fig-6}}

\clearpage

\begin{deluxetable}{rrrrrrrrrr}
\tablecaption{$^{12}$CO $J$ = 2$\rightarrow$1 Integrated Intensity in M17 \label{tbl-1}}
\tablewidth{0pt}
\tablehead{
\colhead{$\Delta\alpha=$} & \colhead{\phs$8^\prime$} & 
\colhead{\phs$6^\prime$}  & \colhead{\phs$4^\prime$} &
\colhead{\phs$2^\prime$}  & \colhead{\phs$0^\prime$} &
\colhead{$-2^\prime$}     & \colhead{$-4^\prime$}    &
\colhead{$-6^\prime$}     & \colhead{$-8^\prime$}
} 
\startdata
$\Delta\delta=16^\prime$ & $<$2.4 & $<$2.4 & 42.4 & 17.8 & 22.5 & 50.8 & 34.3 & 9.5 & $<$2.4 \nl
14\arcmin & 4.5 & $<$2.4 & 58.8 & 37.1 & 35.0 & 6.3 & 29.9 & 13.7 & 8.2 \nl
12\arcmin & 7.5 & 10.4 & 40.0 & 15.7 & 118.6 & 76.1 & 17.4 & 62.8 & 10.3 \nl
10\arcmin & $<$2.4 & 94.5 & 38.6 & 48.3 & 97.4 & 24.1 & 10.5 & 61.6 & $<$2.4 \nl
8\arcmin &  12.2 & 102.6 & 87.3 & 61.9 & 77.8 & 99.9 & 99.1 & 89.1 & 2.9 \nl
6\arcmin &  8.0 & 145.8 & 132.7 & 72.6 & 69.2 & 73.7 & 150.2 & 31.1 & $<$2.4 \nl
4\arcmin &  8.1 & 118.3 & 261.1 & 46.8 & 77.6 & 110.4 & 127.4 & 8.6 & $<$2.4 \nl
2\arcmin &  $<$2.4 & 94.9 & 129.8 & 83.8 & 95.7 & 112.8 & 45.5 & 5.0 & $<$2.4 \nl
0\arcmin &  17.0 & 184.6 & 223.2 & 229.9 & 166.1 & 133.4 & 20.0 & 11.1 & $<$2.4 \nl
$-$2\arcmin &  29.6 & 75.0 & 95.0 & 46.5 & 257.1 & 103.8 & 35.1 & 24.0 & 13.2 \nl
$-$4\arcmin &  114.1 & 20.5 & 134.5 & 29.6 & 239.9 & 142.8 & 98.4 & 39.2 & 12.2 \nl
$-$6\arcmin &  $<$2.4 & 54.0 & 48.0 & 594.0 & 338.1 & 186.5 & 136.3 & 80.1 & $<$2.4 \nl
$-$8\arcmin &  $<$2.4 & 95.8 & 137.0 & 525.3 & 344.4 & 212.9 & 55.0 & 45.0 & 7.8 \nl
$-$10\arcmin &  14.9 & 20.8 & 128.0 & 438.1 & 169.0 & 131.5 & 64.0 & 19.1 & 24.1 \nl
$-$12\arcmin &  13.7 & 37.8 & 21.7 & 145.4 & 138.4 & 98.7 & 77.9 & 55.0 & 81.3 \nl
$-$14\arcmin &  30.5 & 27.9 & 30.7 & 64.5 & 118.6 & 89.4 & 99.5 & 67.2 & 54.1 \nl
$-$16\arcmin  & 16.3 & 15.1 & 15.1 & 35.3 & 16.6 & 72.9 & 76.5 & 152.3 & 103.7 \nl
\enddata
\tablecomments{Units are K km s$^{-1}$ in the hybrid temperature
scale measured in a 22\arcsec\ beam (see \S 2). The formal measurement
uncertainty is typically 0.8 K km s$^{-1}$.  The (0,0) position
corresponds to ($\alpha(1950)$ = 18$^{\rm h}$17$^{\rm m}$21\fs0,
$\delta(1950)$ = $-$16\arcdeg 06\arcmin 45\arcsec). The PDR M17 SW is
located at (2,$-$6).}
\end{deluxetable}

\clearpage

\setlength{\oddsidemargin}{-1.2cm}
\setlength{\evensidemargin}{-0.8cm}
\begin{deluxetable}{rrrrrrrrrr}
\footnotesize
\tablecaption{$^{12}$CO/$^{13}$CO $J$ = 2$\rightarrow$1 Line Ratio in M17 \label{tbl-2}}
\tablewidth{560pt}
\tablehead{
\colhead{$\Delta\alpha=$}     & \colhead{\phs$8^\prime$\pad} & 
\colhead{\phs$6^\prime$\pad}  & \colhead{\phs$4^\prime$\pad} &
\colhead{\phs$2^\prime$\pad}  & \colhead{\phs$0^\prime$\pad} &
\colhead{$-2^\prime$\pad}     & \colhead{$-4^\prime$\pad}    &
\colhead{$-6^\prime$\pad}     & \colhead{$-8^\prime$\pad}
} 
\startdata
$\Delta\delta=16^\prime$ & \nodata\pad & \nodata\pad & 6.8$\pm$0.7 & 10.4$\pm$2.0 & 8.9$\pm$1.2  & 4.2\pad & 12.4$\pm$2.4 & $>$6.0\pad & \nodata\pad \nl
   14\arcmin &  $>$1.5\pad & \nodata\pad & 7.1\pad & $>$15.7\pad & $>$14.0\pad & $>$4.0\pad & 12.6$\pm$2.2  & $>$8.7\pad & $>$5.2\pad \nl
   12\arcmin &  $>$3.8\pad & $>$5.3\pad & 5.7\pad & 6.2$\pm$1.3 & 6.6\pad & 7.6\pad & $>$9.6\pad & 3.0\pad & $>$5.6\pad \nl
   10\arcmin &  \nodata\pad & 3.3\pad & 7.1\pad & 5.4\pad & 4.6\pad & 5.4\pad & 4.0$\pm$0.7  & 3.0\pad & \nodata\pad \nl
    8\arcmin &  3.6$\pm$0.5  & 3.0\pad & 3.1\pad & 4.4\pad & 7.0\pad & 13.3\pad & 5.7\pad & 4.5\pad & $>$2.4\pad \nl
    6\arcmin &  $>$2.4\pad & 3.1\pad & 5.7\pad & 4.9\pad & 3.3\pad & 11.2\pad & 4.7\pad & 8.7$\pm$1.3 & \nodata\pad \nl
    4\arcmin &  $>$3.9\pad & 3.4\pad & 3.2\pad & 4.0\pad & 5.3\pad & 2.9\pad & 3.9\pad & $>$5.4\pad & \nodata\pad \nl
    2\arcmin &  \nodata\pad & 3.9\pad & 5.7\pad & 4.4\pad & 4.5\pad & 3.8\pad & 5.2\pad & $>$3.2\pad & \nodata\pad \nl
    0\arcmin &  $>$8.6\pad & 7.7\pad & 5.9\pad & 4.4\pad & 2.6\pad & 3.7\pad & $>$11.7\pad & $>$4.2\pad & \nodata\pad \nl
 $-$2\arcmin &  $>$13.2\pad & 12.4\pad & 5.9\pad & 7.4\pad & 2.8\pad & 5.7\pad & 9.5$\pm$1.2 & 6.3$\pm$0.9 & $>$8.3\pad \nl
 $-$4\arcmin &  10.4\pad & $>$10.4\pad & 4.3\pad & $>$6.2\pad & 3.7\pad & 4.3\pad & 5.4\pad & $>$13.0\pad & $>$6.6\pad \nl
 $-$6\arcmin &  \nodata\pad & 5.1\pad & 7.4\pad & 2.8\pad & 2.6\pad & 4.1\pad & 7.1\pad & 10.7$\pm$1.1  & \nodata\pad \nl
$-$8\arcmin  &  \nodata\pad & 5.9\pad & 4.5\pad & 2.8\pad & 3.5\pad & 3.3\pad & 6.0\pad & 9.0$\pm$1.0 & $>$4.9\pad \nl
$-$10\arcmin &  $>$5.9\pad & $>$7.5\pad & 4.8\pad & 3.5\pad & 5.0\pad & 6.3\pad & 12.5$\pm$1.3 & $>$6.9\pad & $>$15.2\pad \nl
$-$12\arcmin &  $>$5.2\pad & 6.4$\pm$0.9 & $>$4.7\pad & 6.1\pad & 6.4\pad & 4.1\pad & 4.4\pad & 7.2\pad & 6.3\pad \nl
$-$14\arcmin &  10.1$\pm$1.5 & 21.2$\pm$5.3 & $>$7.8\pad & 7.8\pad & 4.4\pad & 6.5\pad & 3.3\pad & 5.1\pad & 8.2$\pm$1.0 \nl
$-$16\arcmin & $>$4.1\pad & $>$6.4\pad & $>$6.4\pad & 6.0\pad & $>$8.4\pad & 5.5\pad & 10.4\pad & 6.1\pad & 7.4\pad \nl
\enddata
\tablecomments{The line ratio is calculated as the ratio of the
integrated intensity of each line measured in a 22\arcsec\ beam.
Formal measurement uncertainties are given if they exceed 10\%.  The
(0,0) position corresponds to ($\alpha(1950)$ = 18$^{\rm h}$17$^{\rm m}$
21\fs0, $\delta(1950)$ = $-$16\arcdeg 06\arcmin 45\arcsec). The PDR
M17 SW is located at (2,$-$6).}
\end{deluxetable}

\clearpage

\setlength{\oddsidemargin}{0.0cm}
\setlength{\evensidemargin}{0.0cm}
\begin{deluxetable}{cccccccccc}
\tablecaption{$^{12}$CO and $^{13}$CO $J$ = 3$\rightarrow$2 Lines in M17 \label{tbl-3}}
\tablewidth{0pt}
\tablehead{
\colhead{$\Delta\alpha$} & \colhead{$\Delta\delta$} &
\colhead{$I(\rm^{12}CO)$} & 
\colhead{\underline{$^{12}$CO(3$\rightarrow$2)}} & \colhead{} &\colhead{}  &
\colhead{$\Delta\alpha$} & \colhead{$\Delta\delta$} & 
\colhead{$I(\rm^{12}CO)$} & 
\colhead{\underline{$^{12}$CO(3$\rightarrow$2)}}  \\
\colhead{(\arcmin)} & \colhead{(\arcmin)}& 
\colhead{(K km s$^{-1}$)} & \colhead{$^{13}$CO(3$\rightarrow$2)} & \colhead{} & \colhead{} &
\colhead{(\arcmin)} & \colhead{(\arcmin)}& 
\colhead{(K km s$^{-1}$)} & \colhead{$^{13}$CO(3$\rightarrow$2)}
} 
\startdata
\phs8 & 8 &  $<$4.8    &     \nodata     & & &  $-$8 & \phs\phn0 &   $<$4.8   &    \nodata      \nl 
\phs6 & 8 &  \phn76.7   & \phs$1.8\pm0.1$ & & & \phs8 &  \phn$-$8 &   $<$4.8   &    \nodata      \nl
\phs4 & 8 &  \phn58.7   & \phs$2.0\pm0.1$ & & & \phs6 &  \phn$-$8 &   \phn82.5  & \phs$7.7\pm0.4$ \nl
\phs2 & 8 &  \phn39.5   & \phs$4.1\pm0.5$ & & & \phs4 &  \phn$-$8 & 114\phm{.0} & \phs$5.4\pm0.2$ \nl
\phs0 & 8 &  \phn57.0   & \phs$7.4\pm1.1$ & & & \phs2 &  \phn$-$8 & 404\phm{.0} & \phs$2.1\pm0.1$ \nl
 $-$2 & 8 &  \phn58.3   &  $>$$7.1\pm1.6$ & & & \phs0 &  \phn$-$8 & 250\phm{.0} & \phs$2.9\pm0.1$ \nl
 $-$4 & 8 &  \phn74.7   & \phs$3.0\pm0.2$ & & &  $-$2 &  \phn$-$8 & 153\phm{.0} & \phs$2.6\pm0.1$ \nl
 $-$6 & 8 &  \phn68.2   & \phs$2.9\pm0.1$ & & &  $-$4 &  \phn$-$8 &   \phn26.5  & \phs$5.0\pm0.7$ \nl
 $-$8 & 8 &  $<$4.8    &     \nodata     & & &  $-$6 &  \phn$-$8 &   \phn27.2  &  $>$$7.4\pm2.1$ \nl
\phs8 & 0 & \phn\phn7.7 &  $>$$2.7\pm0.6$ & & &  $-$8 &  \phn$-$8 &   $<$4.8   &    \nodata      \nl
\phs6 & 0 & 126\phm{.0} & \phs$5.8\pm0.2$ & & & \phs2 &    \phs16 &   \phn16.2  & \phs$6.5\pm1.3$ \nl
\phs4 & 0 & 184\phm{.0} & \phs$3.9\pm0.1$ & & & \phs2 &    \phs12 & \phn\phn7.5 & \phs$2.1\pm0.5$ \nl
\phs2 & 0 & 155\phm{.0} & \phs$2.3\pm0.1$ & & & \phs2 & \phs\phn6 &   \phn49.3  & \phs$2.6\pm0.2$ \nl
\phs0 & 0 & 119\phm{.0} & \phs$2.0\pm0.1$ & & & \phs2 &  \phn$-$6 & 440\phm{.0} & \phs$1.4\pm0.1$ \nl
 $-$2 & 0 &  \phn92.8   & \phs$2.8\pm0.1$ & & & \phs2 &     $-$12 & 105\phm{.0} & \phs$3.6\pm0.7$ \nl
 $-$4 & 0 & \phn\phn8.5 &  $>$$5.1\pm1.6$ & & & \phs2 &     $-$16 &   \phn21.3  & \phs$5.6\pm0.6$ \nl
 $-$6 & 0 &  $<$4.8    &     \nodata     & & &       &           &             &                 \nl
\enddata
\tablecomments{The line ratio is calculated as the ratio of the
integrated intensity of each line measured in a 15\arcsec\ beam.  The
typical measurement uncertainty in the $^{12}$CO integrated intensity
is 1.6 K km s$^{-1}$.  The (0,0) position corresponds to
($\alpha(1950)$ = 18$^{\rm h}$17$^{\rm m}$21\fs0, $\delta(1950)$ =
$-$16\arcdeg 06\arcmin 45\arcsec).  The PDR M17 SW is located at
(2,$-$6).}
\end{deluxetable}

\clearpage

\setlength{\oddsidemargin}{-1.0cm}
\setlength{\evensidemargin}{-1.0cm}
\begin{deluxetable}{cccccccc}
\tablecaption{$J$ = 2$\rightarrow$1 and $J$ = 3$\rightarrow$2 Line Ratios at Nine Positions in M17 \label{tbl-4}}
\tablewidth{0pt}
\tablehead{
\colhead{$\Delta\alpha$} & \colhead{$\Delta\delta$}& 
\colhead{$I$[$^{12}$CO(3$\rightarrow$2)]}& 
\colhead{$I$[$^{13}$CO(3$\rightarrow$2)]}  & \colhead{\underline{$^{12}$CO(3$\rightarrow$2)}} & 
\colhead{\underline{$^{12}$CO(2$\rightarrow$1)}} 
& \colhead{\underline{$^{12}$CO(3$\rightarrow$2)}}& 
\colhead{\underline{$^{13}$CO(3$\rightarrow$2)}} \\
\colhead{(\arcmin)} & \colhead{(\arcmin)}& 
\colhead{(K km s$^{-1}$)}& 
\colhead{(K km s$^{-1}$)}    & \colhead{$^{12}$CO(2$\rightarrow$1)} & 
\colhead{$^{13}$CO(2$\rightarrow$1)} & \colhead{$^{13}$CO(3$\rightarrow$2)}& 
\colhead{$^{13}$CO(2$\rightarrow$1)} 
} 
\startdata
\phs4 & \phs\phn4 & 205\phd\phn & 115\phd\phn & 0.78 & 3.2 & 1.8 & 1.39 \nl
\phs2 &    \phs12 &  \phn11.5   & \phn\phn3.2 & 0.74 & $6.2\pm 1.3$ & 3.6 & $1.27\pm 0.28$ \nl
\phs2 & \phs\phn6 &  \phn45.5   &  \phn17.5   & 0.63 & 4.9 & 2.6 & 1.19 \nl
\phs2 & \phs\phn0 & 148\phd\phn &  \phn63.3   & 0.66 & 4.4 & 2.3 & 1.22 \nl
\phs2 &  \phn$-$6 & 492\phd\phn & 312\phd\phn & 0.83 & 2.8 & 1.6 & 1.45 \nl
\phs2 &     $-$12 & 106\phd\phn &  \phn28.8   & 0.73 & 6.1 & 3.7 & 1.21 \nl
\phs0 &    \phs12 &  \phn94.2   &  \phn18.8   & 0.80 & 6.6 & 5.0 & 1.04 \nl
 $-$4 & \phs\phn6 & 105\phd\phn &  \phn35.3   & 0.69 & 4.7 & 3.0 & 1.09 \nl
 $-$4 &     $-$14 &  \phn64.5   &  \phn23.0   & 0.66 & 3.3 & 2.8 & 0.78 \nl
\enddata
\tablecomments{The line ratio is calculated as the ratio of the
integrated intensity of each line measured in a 22\arcsec\ beam.
Measurement uncertainties are noted
where they exceed 10\%. Typical measurement uncertainties for
the integrated intensities are 0.8 K km s$^{-1}$ for
$^{12}$CO $J$ = 2$\rightarrow$1, 0.7 K km s$^{-1}$ for
$^{13}$CO $J$ = 2$\rightarrow$1, 0.5 K km s$^{-1}$ for
$^{12}$CO $J$ = 3$\rightarrow$2, and 0.3 K km s$^{-1}$ for
$^{13}$CO $J$ = 3$\rightarrow$2.
The (0,0) position corresponds to
($\alpha(1950)$ = 18$^{\rm h}$17$^{\rm m}$21\fs0,
$\delta(1950)$ = $-$16\arcdeg 06\arcmin 45\arcsec). 
The PDR M17 SW is located at (2,$-$6).}
\end{deluxetable}

\begin{table}
\dummytable\label{tbl-5}
\end{table}

\begin{table}
\dummytable\label{tbl-6}
\end{table}

\end{document}


\setcounter{table}{4}

\setlength{\oddsidemargin}{1.5cm}
\setlength{\evensidemargin}{1.5cm}
\begin{table}
\caption{Comparison of Line Ratios with Previous Galactic Observations
\label{tbl-6}}
\begin{tabular*}{480pt}{@{\extracolsep{\fill}}lccc}
\tablecolumns{4}
 & \multicolumn{2}{c}{} & \nl
  \multicolumn{4}{c}{A. $^{12}$CO ($J$=3$\rightarrow$2)/($J$=2$\rightarrow$1) Line Ratio} \nl
\tableline
\tableline
\colhead{Cloud} & \colhead{Ratio} & 
\colhead{Ref.} & \colhead{Comments} \nl
\tableline
M17 & $0.76\pm0.19$ & 1 & $16^\prime\times 32^\prime$ \nl
M17 & 0.7-1.4\tablenotemark{a} & 2 & PDR \nl
Orion B & $\sim$1 & 3 & \nl
Rosette & $\sim$1 & 4 & \nl
Galactic Plane & $0.55\pm0.08$ & 5 & pencil beam survey \nl
Cepheus & 0.45 & 6 & isolated core \nl
Taurus,CygOB7 & $0.55\pm0.1$ & 7 & cloud edges \nl
high latitude clouds & 0.2-0.7\tablenotemark{a} & 8 & may decrease towards edge \nl
dark globules & 0.5-0.7 & 9 & \nl
\tableline
\tablecolumns{4}
 & \multicolumn{2}{c}{} & \nl
  \multicolumn{4}{c}{B. $^{12}$CO/$^{13}$CO $J$=2$\rightarrow$1 Line Ratio} \nl
\tableline
\tableline
\colhead{Cloud} & \colhead{Ratio} & 
\colhead{Ref.} & \colhead{Comments} \nl
\tableline
M17 & $4.5\pm 0.7$\tablenotemark{b} & 1 & $16^\prime\times 32^\prime$ \nl
M17 & 1-2 & 12 & near PDR \nl
Orion A & 2 & 13 & main ridge \nl
S140 & 2.4 & 10 & PDR peak \nl
Galactic Plane & $5.5\pm1$ & 5 & pencil beam survey \nl
Orion A and B & 4-6\tablenotemark{b} & 14 & large-area map \nl
Cepheus & 5.1\tablenotemark{b} & 6 & 22 positions \nl
\tableline
\tablecolumns{4}
 & \multicolumn{2}{c}{} & \nl
  \multicolumn{4}{c}{C. $^{12}$CO/$^{13}$CO $J$=3$\rightarrow$2 Line Ratio} \nl
\tableline
\tableline
\colhead{Cloud} & \colhead{Ratio} & 
\colhead{Ref.} & \colhead{Comments} \nl
\tableline
M17 & $3.7\pm0.9$ & 1 & $16^\prime\times 32^\prime$ \nl
M17 & 1-1.7 & 11 & $2\times2^\prime$; 21 km s$^{-1}$ component \nl
M17 & 1.4-$>$10 & 11 & 14 and 26 km s$^{-1}$ components \nl
Orion B & $\sim$1 & 3 & strip map \nl
Rosette & $\sim$1 & 4 & two positions\nl
S140 & 1.5 & 10 & PDR peak \nl
Galactic Plane & $\sim$13 & 5 & pencil beam survey \nl
\tableline
\tablecolumns{4}
 & \multicolumn{2}{c}{} & \nl
  \multicolumn{4}{c}{D. $^{13}$CO ($J$=3$\rightarrow$2)/($J$=2$\rightarrow$1) Line Ratio} \nl
\tableline
\tableline
\colhead{Cloud} & \colhead{Ratio} & 
\colhead{Ref.} & \colhead{Comments} \nl
\tableline
M17 & $1.3\pm0.3$ & 1 & $16^\prime\times 32^\prime$ \nl
S140 & 1.1\tablenotemark{a} & 10 & PDR peak \nl
Galactic Plane & 0.23\tablenotemark{c} & 5 & pencil beam survey \nl
\tableline
\end{tabular*}
\vspace{-19.8pt}
\tablenotetext{}{\normalsize References: 1. This paper. 2. Stutzki et al.
1988. 3. Kramer  et al. 1996. 4. Schneider et al. 1998. \\ 5. Sanders et al. 1993.
6. Heyer \& Ladd 1995. 7. Falgarone et al. 1991. 8. van Dishoeck et 
al. 1991. \\
9. Avery et al. 1987. 10. Minchin et al. 1993. 11. Greaves et al. 1992. 
12.  Stutzki \& G\"usten 1990. \\ 13. Castets et al. 1990. 14. Sakamoto et al.
1994.}
\tablenotetext{a}{\normalsize Assumes 
 $^{12}$CO ($J$=2$\rightarrow$1)/($J$=1$\rightarrow$0) = 0.7.}
\tablenotetext{b}{\normalsize Line ratios increase towards
the outskirts of the cloud.}
\tablenotetext{c}{\normalsize This ratio is unusually small due to
the unusually large value of the $^{12}$CO/$^{13}$CO $J$=3$\rightarrow$2
ratio.}
\end{table}